\documentclass[twocolumn]{aastex63}

\hypersetup{linkcolor=red, citecolor=blue}
\usepackage{comment}
\usepackage{amssymb}
\usepackage{amsfonts}
\usepackage{amsmath}
\usepackage[varg]{txfonts}
\usepackage{natbib}
\usepackage{color}
\usepackage{graphicx}
\usepackage{hyperref}

\newcommand{\RNum}[1]{\uppercase\expandafter{\romannumeral #1\relax}}
\newcommand{\felix}[1]{\textsc{F\small{ELIX}}}

\received{---}
\revised{---}
\accepted{---}
\submitjournal{ApJ}

\shorttitle{An apparently cool DBV from K2}
\shortauthors{Duan et al. 2021}
\begin{document}

\title{EPIC 228782059: Asteroseismology of what could be the coolest pulsating helium-atmosphere white dwarf (DBV) known?}

\email{weikai.zong@bnu.edu.cn}

\author{R.M.~Duan}
\affiliation{Department of Astronomy, Beijing Normal University, Beijing 100875, P.~R.~China}

\author[0000-0002-7660-9803]{W.~Zong}
\affiliation{Department of Astronomy, Beijing Normal University, Beijing 100875, P.~R.~China}

\author[0000-0001-8241-1740]{J.-N.~Fu}
\affiliation{Department of Astronomy, Beijing Normal University, Beijing 100875, P.~R.~China}

\author{Y.H.~Chen}
\affiliation{Institute of Astrophysics, Chuxiong Normal University, Chuxiong 675000, P.~R.~China}
%\affiliation{AAS Journals Associate Editor-in-Chief}

\author[0000-0001-5941-2286]{J.~J.~Hermes}
\affiliation{Department of Astronomy $\&$ Institute for Astrophysical Research, Boston University, 725 Commonwealth Ave., Boston, MA 02215, USA}
%\affiliation{AAS Journals Associate Editor-in-Chief}

\author[0000-0002-0853-3464]{Zachary~P.~Vanderbosch}
\affiliation{Department of Astronomy, The University of Texas at Austin, Austin, TX 78712, USA}
%\affiliation{AAS Journals Associate Editor-in-Chief}

\author{X.Y.~Ma}
%\affiliation{AAS Director of Publishing}
\affiliation{Department of Astronomy, Beijing Normal University, Beijing 100875, P.~R.~China}

\author[0000-0002-6018-6180]{S.Charpinet}
\affiliation{Institut de Recherche en Astrophysique et Plan\'{e}tologie,~CNRS,~Universit\'{e} de Toulouse, CNES,~14 Avenue Edouard Belin,~31400 Toulouse,~France}

%% Note that the \and command from previous versions of AASTeX is now
%% depreciated in this version as it is no longer necessary. AASTeX
%% automatically takes care of all commas and "and"s between authors names.

%% AASTeX 6.1 has the new \collaboration and \nocollaboration commands to
%% provide the collaboration status of a group of authors. These commands
%% can be used either before or after the list of corresponding authors. The
%% argument for \collaboration is the collaboration identifier. Authors are
%% encouraged to surround collaboration identifiers with ()s. The
%% \nocollaboration command takes no argument and exists to indicate that
%% the nearby authors are not part of surrounding collaborations.
%%\Large
%% Mark off the abstract in the ``abstract'' environment.
\begin{abstract}
We present analysis of a new pulsating helium-atmosphere (DB) white dwarf, EPIC~228782059, discovered from 55.1~days of {\em K2} photometry. The long duration, high quality light curves reveal 11 independent dipole and quadruple modes, from which we derive a rotational period of $34.1 \pm 0.4$~hr for the star. An optimal model is obtained from a series of grids constructed using the White Dwarf Evolution Code, which returns $M_{*} = 0.685 \pm 0.003 M_{\odot}$, $T_{\rm{eff}}= 21{,}910 \pm 23$\,K and $\log g = 8.14 \pm0.01$\,dex. These values are comparable to those derived from spectroscopy by Koester \& Kepler ($20{,}860 \pm 160$\,K and $7.94 \pm0.03$\,dex). If these values are confirmed or better constrained by other independent works, it would make EPIC~228782059 one of the coolest pulsating DB white dwarf star known, and would be helpful to test different physical treatments of convection, and to further investigate the theoretical instability strip of DB white dwarf stars.

\end{abstract}

%% Keywords should appear after the \end{abstract} command. 
%% See the online documentation for the full list of available subject
%% keywords and the rules for their use.
\keywords{asteroseismology-star: individual: EPIC228782059, DB type, white dwarf}

%% From the front matter, we move on to the body of the paper.
%% Sections are demarcated by \section and \subsection, respectively.
%% Observe the use of the LaTeX \label
%% command after the \subsection to give a symbolic KEY to the
%% subsection for cross-referencing in a \ref command.
%% You can use LaTeX's \ref and \label commands to keep track of
%% cross-references to sections, equations, tables, and figures.
%% That way, if you change the order of any elements, LaTeX will
%% automatically renumber them.

%% We recommend that authors also use the natbib \citep
%% and \citet commands to identify citations.  The citations are
%% tied to the reference list via symbolic KEYs. The KEY corresponds
%% to the KEY in the \bibitem in the reference list below.
\section{Introduction} %\label{sec:intro}

White dwarfs mark the graveyard and evolutionary destiny for all low-mass stars when nuclear burning ceases fire in their cores. About 98\% of stars in our Galaxy will evolve into white dwarfs \citep{2008ARA&A..46..157W},
including our Sun. These faint blue stars offer extreme conditions to test fundamental physics that is far from reach by the current terrestrial laboratory \citep[see, e.g.,][]{2020Natur.584...51K,2018PhRvD..98j3023N}. They can also be the tracer to ages for cosmology and Galactic archaeology \citep{2001PASP..113..409F}.

When white dwarfs evolve along cooling tracks, they go through several different regions where they pulsate, mainly the DOV, DBV and DAV instability strips. As they cool, their brightness can be modulated as the result of convection-driven pulsations \citep[see, e.g.,][and reference therein]{2008ARA&A..46..157W,2008PASP..120.1043F,2010A&ARv..18..471A}. 

Among all types of pulsating white dwarfs, the DBV (or V777 Her) class refers to those with helium-dominated atmospheres, ranging from 
roughly $22{,}000 < T_{\rm{eff}} < 32{,}000$\,K and $7.5 < \log g < 8.3$\,dex \citep{2019A&ARv..27....7C}. The first DBV was discovered immediately following the theoretical prediction of its existence by \citet{1982ApJ...252L..65W}. We now know of roughly 40 DBV stars, which is about 11\% of all pulsating white dwarfs known \citep{2019A&ARv..27....7C,2018whitedwarf..1025c.}.

The richness of pulsations in non-radial $g-$modes  offers a new window to probe the interior structure and chemical profiles of a DBV star \citep[see, e.g.,][]{1982ApJ...259..219R}. In order to produce better frequency resolution in pulsation spectra for DBV observations, a global network called the Whole Earth Telescope \citep[see, e.g.,][]{1990ApJ...361..309N} was organized in order to collect photometric data for several particular interesting DBVs, such as GD~358 \citep[see, e.g.,][]{2003BaltA..12...45K}. With its numerous pulsations observed, GD~358 has been explored for various physical tests of convection, magnetism, and rotation \citep{2009ApJ...693..564P,2010ApJ...716...84M}. The long-term observations of DBV stars may also provide constraints of the cooling rate of hot white dwarfs, where neutrinos contribute meaningfully to the luminosity \citep{2004ApJ...602L.109W}.

The {\em Kepler} mission delivered unprecedented, high-quality photometric data, leading to various breakthroughs in stellar physics, in particular, probing the internal physical processes of pulsating stars via the technique of asteroseismology  \citep[see, e.g.,][]{2018Natur.554...73G,2020FrASS...7...47C}. There was only one DBV star, KIC~08626021, discovered with multiple pulsations in the original {\em Kepler} field \citep{2011ApJ...736L..39O}. Due to its precise frequency resolution, it has been the subject of intensive studies to improve theoretical models, such as those of the parametric evolutionary models of the White Dwarf Evolution Code  \citep[\texttt{WDEC};][]{2018AJ....155..187B} and the fully evolutionary models of the La Plata code \citep[\texttt{LPCODE};][]{2013ASPC..469...41C}, where the latter harbors chemical structures resulting from the complete evolutionary history of the progenitor stars.

With further observations by {\em Kepler}, \citet{2014ApJ...794...39B} identified seven independent modes to re-construct the \texttt{WDEC} model for KIC~08626021, revealing a very thin helium layer and hot temperature. Additionally, \citet{2016A&A...585A..22Z} discovered clear amplitude and frequency modulations occurring in several multiplets in that star using the same data. They attribute those modulations to nonlinear mode interactions, which lead to identification of one additional independent mode. With eight modes, \citet{2018Natur.554...73G} performed a static forward model to KIC~08626021 and obtained an unprecedentedly precise solution, measuring a large core with a higher oxygen-to-carbon ratio than previous solutions. The results were challenged by \citet{2018ApJ...867L..30T}, who suggested that neutrino cooling would also impact the seismic solution to KIC~08626021 at such a high effective temperature. A quick response by \citet{2019A&A...628L...2C} suggested that the neutrino emission, indeed, should be incorporated in the static models, but does not have a significant effects on the seismic properties derived for KIC~08626021.  Nevertheless,~\citet{2019A&A...630A.100D} conclude that the current white dwarf formation and evolution can hardly reproduce the internal chemical results of KIC~08626021 derived by \citet{2018Natur.554...73G}.

After a failure of the second reaction wheel controlling pointing of the spacecraft, {\em Kepler} began observations of new campaigns along the ecliptic plane \citep{2014PASP..126..398H}. This strategy, also called the {\em K2} mission, produced more white dwarfs to be monitored, enabling observations of many more pulsating white dwarfs \citep{2017ApJS..232...23H}. To date, only one DBV observed by {\em K2} has been published. That DBV, PG~0112+104, has stable pulsation frequencies and a rotation period of 10.2\,hr \citep{2017ApJ...835..277H}.

Here we report the discovery of a new DBV star, EPIC~228782059, located near the cool edge of the DBV instability strip. The paper is organised as follows:  In Section~2 we describe {\em K2} photometry and frequency content for EPIC~228782059. We present the seismic results from the \texttt{WDEC} code and discuss its location in the DBV instability strip in Section~3. The conclusion is given in the final section.

\section{Frequency content}

\label{sec:style}

\subsection{Photometry}
\begin{figure*}
    \centering 
    \includegraphics[width=\textwidth]{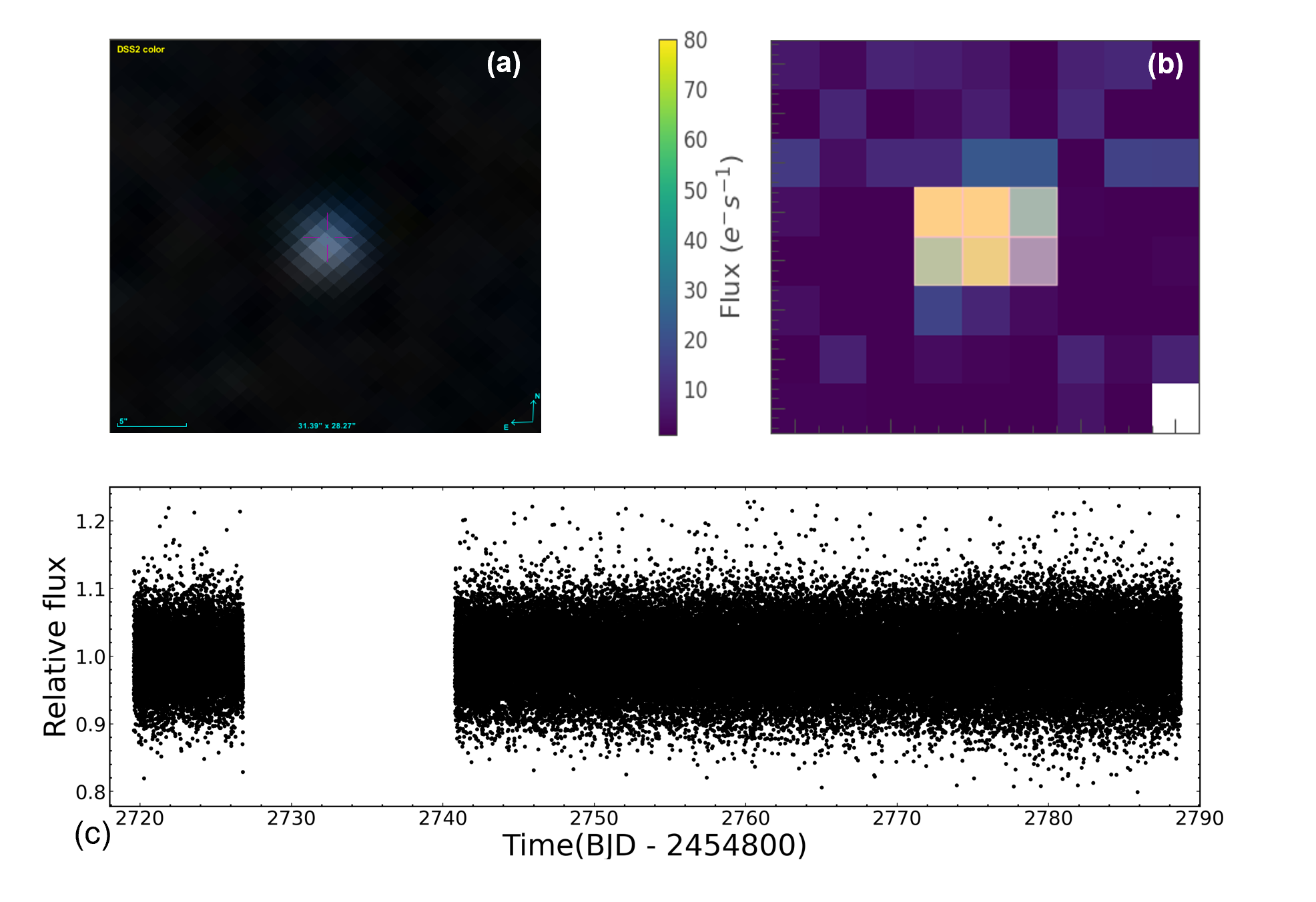}
    \caption{(a) The finder chart of EPIC~228782059 from Aladin~and (b) the image from the {\em Kepler} target pixel file with the optimal aperture marked.  (c) The relative flux of {\em K2} photometry for EPIC~228782059, with motion-correlated systematic trends removed.}
    \label{fig:lightcurve_bin.eps}
\end{figure*}

EPIC~228782059 (SDSS~J124538.21-073138.3, $K_\mathrm{p}$ = 18.4\,mag, Figure\,\ref{fig:lightcurve_bin.eps}a) was identified as a DB-type white dwarf from the Sloan Digital Sky Survey (SDSS) by \citet{2013ApJS..204....5K}. An analysis of atmospheric parameters of DBs from the SDSS found for this star $T_\mathrm{eff}=20,868\pm160$~K and $\log g = 7.914 \pm0.027$~dex \citep{2015A&A...583A..86K}. 
EPIC~228782059 was proposed to be observed by {\em K2} in Campaign~10. The target pixel files (TPFs) in short-cadence (58.85~s) were downloaded from the Mikulski Archive for Space Telescopes, and we use the \texttt{Lightkurve} \citep{2018ascl.soft12013L} package to extract photometry from the TPFs. As {\em K2} suffers a $\sim6.5$~hr thruster firing to compensate for the solar pressure variation for fine pointing, we subsequently used the \texttt{KEPSFF} routine \citep{2014PASP..126..948V} to correct the systematic photometric variation that is induced by the low-frequency motion of the target on the CCD module. A series of apertures of different pixel sizes were tested on the TPF to optimise the photometry. We finally chose a fixed six-pixel aperture to extract our light curve, as shown in Figure\,\ref{fig:lightcurve_bin.eps}(b).

After extracting the photometry we fit out a third-order polynomial and sigma clipped the light curve to 4.5$\sigma$ in order to detrend the light curve and to clip the outliers, leaving $79{,}378$ data points. Figure\,\ref{fig:lightcurve_bin.eps}(c) shows the final light curve of EPIC~228782059 during C10 with a duration of 55.1\,days. We note that there is a large gap about a week into the light curve due to the failure of CCD Module 4 during {\em K2} Campaign 10.

\subsection{Frequency analysis}
\begin{table*}
	\centering
	\caption{The frequency content detected from {\sl K}2 photometry and their mode identification in EPIC~228782059.}
	\label{tab:1_table}
	\normalsize
	\begin{tabular}{lccccrrcrcrc}
		\hline
		ID & Freq. &$\sigma$ Freq. & $\rm P_{\mathrm{obs}}$ & $\rm P_{\rm{cal}}$ & $\rm P_{\rm{obs}} -\rm P_{\rm{cal}}$& Amplitude & $\sigma$ \rm{Amp}& SNR &$\ell$&$m$&$n$\\
		&($\mu$Hz)&($\mu$Hz)&(s)&(s)&(s)~~~~~~&(ppt)~~~~&(ppt) & & & & \\
		\hline
		% &  & & { Independent~Pulsation~Modes} & & & & \\
        $f_{1,0^\dagger}$&2922.545&0.002 &342.167&342.388& -0.221&13.05&0.27&48.7&1&0&6\\
        $f_{{1,+1^*}}$&2926.592&0.030&341.694& & &0.81&0.27 &3.1&1&1&...\\
        $f_{{2,0^\dagger}}$&2927.373&0.009 &341.603&341.460&0.143&2.78&0.26&10.4&2&0&12\\
        $f_{{3,-1}}$&3428.662&0.010&291.658& & &2.31&0.26&8.9&1& -1&...\\
        $f_{3,0}$&3432.711&0.019&291.314&291.345& -0.031&1.26&0.26&4.9&1&0&5\\
        $f_{{3,+1}}$&3436.662&0.005&290.979& & &4.95&0.26&19.0&1&1&...\\
        $f_{4,-1}$&3773.867&0.016&264.980& & &1.52&0.26&5.8&1& -1&...\\
        $f_{{4,0}}$&3777.989&0.018&264.691&264.645&0.046&1.33&0.26&5.0&1&0&4\\
        $f_{{4,+1^*}}$&3782.053&0.026&264.406& & &0.93&0.26&3.5&1&1&...\\
        $f_{{5},-2}$&4345.170&0.009&230.140& & &2.66&0.26&10.1&2& -2&...\\
        $f_{{5,-1}}$&4351.986&0.013&229.780& & &1.88&0.26&7.1&2& -1&...\\
        $f_{5,0}$&4358.875&0.011&229.416&229.532& 0.116&2.20&0.26&8.0&2&0&7\\
        $f_{{5,+2}}$&4372.535&0.009&228.700& & &2.79&0.26&10.6&2&2&...\\
		\hline
		$\chi$ &  & & & &0.131& &  & & & & \\
		\hline
		~~~~~~Linear & Combinations & Frequencies &    & & & & \\
		$f_{{1,0}}$ + $f_{{3,+1}}$&6369.216&0.017&157.252&&&1.43&0.27&5.4&...&...&...\\
		$f_{{5,-2}}$  -  $f_{1,0}$& 1422.572&0.013&702.952& & & 1.92& 0.26&7.4&...&...&...\\
		\hline
	\end{tabular}
    \tablenotetext{\dagger}{The $m = 0$ identified with the minimum frequency difference between the observation and seismic models.} \tablenotetext{*}{Frequencies below the 4.8$\sigma$ detection limits but consistent with the mode identification from seismic models.}
\end{table*}

\begin{figure*}
	\centering 
	\includegraphics[width=\textwidth]{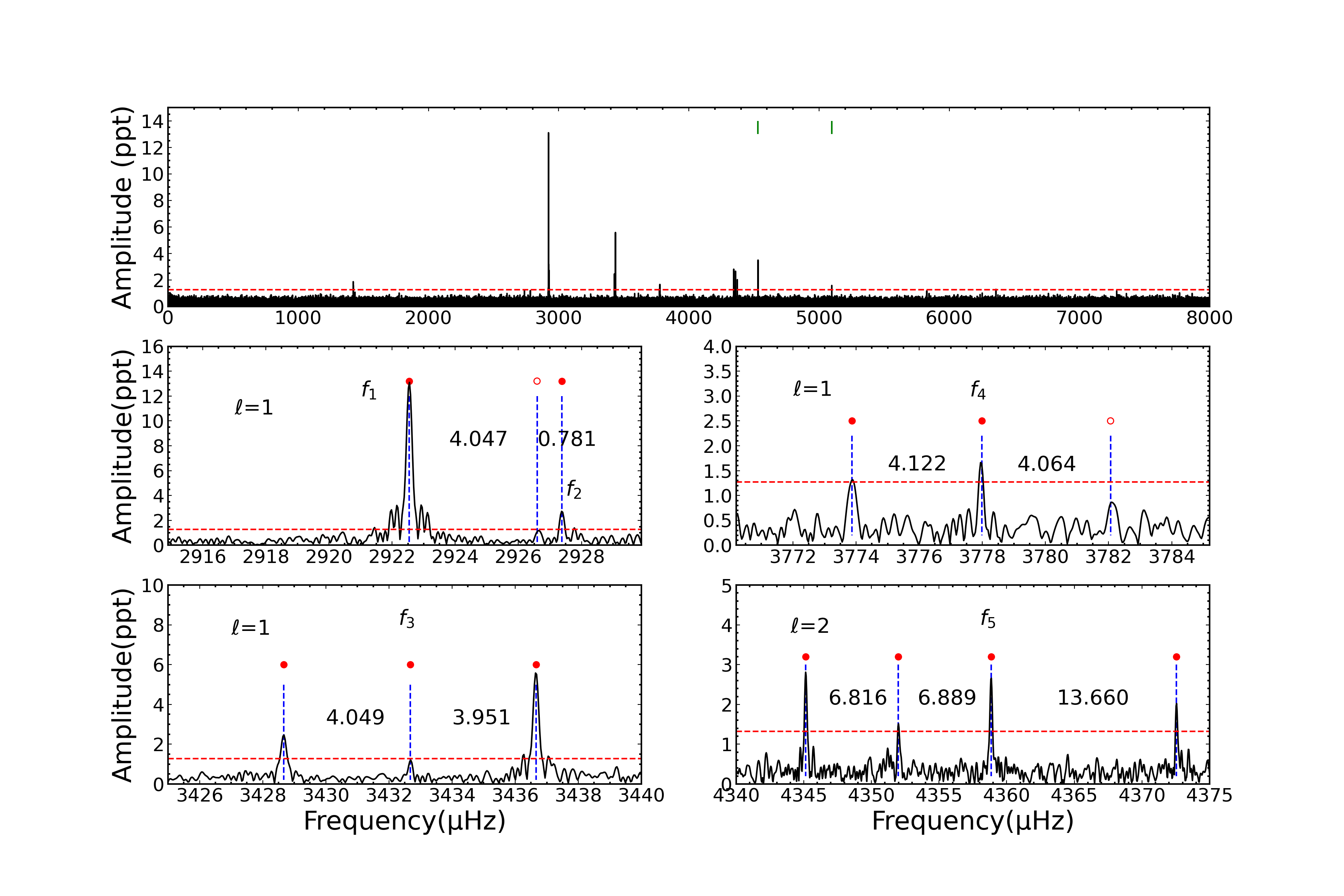}
	\caption{The Lomb-Scargle periodogram of EPIC~228782059. The horizontal lines (in red) denote the 4.8$\sigma$ significance threshold. Two vertical short segments indicate where the instrumental artifacts at integer harmonics of the long-cadence exposure time are identified (top panel). The solid points represent the frequencies detected above the 4.8$\sigma$ threshold, whereas the open circles are lower than 4.8$\sigma$ but probably the real ones indicated by their frequency spacing (number in lower panels). The modal degree numbers are given by their frequency spacing values.}
	\label{fig:Modes_bin.eps}
\end{figure*}

The flexible software package \texttt{FELIX}\footnote{Frequency Extraction for Lightcurve exploitation, which was developed by St\'{e}phane~Charpinet. See details of the code in \citet{2010A&A...516L...6C}.} was used for Fourier decomposition of the light curve, which incorporates the standard prewhitening and nonlinear least-square fitting techniques for Fourier transformation \citep{1975Ap&SS..36..137D,1982ApJ...263..835S}. The threshold of acceptance for a significant frequency is set to be 4.8$\sigma$ of the local noise level, tuned from previous work with {\em Kepler} and {\em TESS} \citep{2016A&A...585A..22Z,2019A&A...632A..90C}. Table~\ref{tab:1_table} lists the frequencies with their attributes extracted with \texttt{Felix}: the ID ordered by their frequency (Column~1), frequency (Column~2), amplitude (Column~7) and their uncertainties (Column~3 and 8), the corresponding period (Column~4), the signal-to-noise ratio (SNR, Column~9), and the other quantities derived from seismic models.

Figure\,\ref{fig:Modes_bin.eps} shows the Lomb-Scargle periodogram of EPIC~228782059: significant signals are found from roughly $1500$--$6300$\,$\mu$Hz. There are 11 frequencies detected and two additional linear combination frequencies. We note that two frequencies detected are below the significance threshold but their positions correspond well to equally spaced rotational components. The narrow profile of the rotational multiples~$f_{1,0}=2922.5$\,$\mu$Hz, $f_{3,0}=3432.7$\,$\mu$Hz, $f_{4,0}=3777.9$\,$\mu$Hz, $f_{5,0}=4358.8$\,$\mu$Hz are shown in the lower panels. Two values of equal frequency spacing are identified, i.e., $\approx$4.0\,$\mu$Hz and $\approx$6.8\,$\mu$Hz for $\ell = 1$ and $\ell = 2$, respectively, according to the rotational formulae that is truncated to the first order for slow  rotation \citep[see, e.g.,][]{1992ApJ...394..670D},
\begin{equation}
\omega_{nlm}=\omega_{nl} +
m(1-C_{nl})\Omega.
\end{equation}
Here $\omega_{nlm}$ is the rotational splitting frequency, $\omega_{nl}$ is the non-rotational split frequency, $\Omega$ is the rotational frequency, and the Ledoux constant $C_{nl} \approx 1/ \ell(\ell +1)$ for high radial-order $g$-modes.

\subsection{Astroseismological analysis}
Mode identification, obtained from observational constraints, is crucial to determine the seismic solutions for various theoretical models. As indicated from Equation~(1), the rotational splitting frequencies provide a conventional way to identify the spherical harmonic degree $\ell$ and the azimuthal order $m$. For $g$-mode pulsations in chemically homogeneous and radiative white dwarfs, the consecutive, higher radial orders ($n\gg \ell$) follow a pattern of equal period spacing, following approximately the asymptotic regime \citep{1990ApJS...72..335T}, which depends only on the structure of a model:
\begin{equation}
	\Delta \Pi_{l}\approx  \frac{\Pi_{0}}{\sqrt{(l(l+1))}} %\epsilon \label{eq1}
\end{equation}
with $\Pi_{0}$ defined as,
\begin{equation}
     \Pi_{0} = 2 \pi^2(\int_{1}^{R} \frac{N}{r}dr)^{-1}
\end{equation}
where $N$ is the Brunt-V\"ais\"al\"a~frequency and $r$ is the radial coordinate.

A complete triplet is found at 3432~$\mu$Hz with three components with frequency spacing of $\Delta f \sim 4.0~\mu$Hz. Another triplet is resolved at 3778~$\mu$Hz with $\Delta f \sim 4.1~\mu$Hz, but the prograde component ($m = +1$) is below the detection threshold. A doublet is identified at 2922.5~$\mu$Hz with a suspected prograde component whose $m$ could determined by a nearby peak significantly found at 2927.3~$\mu$Hz. An incomplete quintuplet is discovered near 4560~$\mu$Hz, with four significant components, but the $m=+1$ one is missing, with a frequency spacing of $\Delta f \approx 6.8~\mu$Hz. With those identified frequency splittings, we estimate a rotation period of $34.1\pm0.4$~hr for EPIC~228782059.

From the above triplets with periods at 342.2~s, 291.3~s and 264.7~s, we calculate a preliminary period spacing $\Delta\,\Pi_{1}=25.7\,s$ for $\ell = 1$ modes  from the linear relationship of $\Pi_{1,n} = 25.7 \times n + 239.4$. The $\ell = 2$ modes have a period spacing $\Delta\,\Pi_2 = 14.8$~s, derived from $\Delta\,\Pi_{2}=1/\sqrt{3}\Delta\Pi_{1}$. However, this relation is only accurate for high-order modes typically with long period, $P>500$~s \citep{1992ApJS...81..747B}. Our preliminary calculation suggests that the mode at 2927\,$\mu$Hz is very possibly identified as $\ell = 2$, since it is very close to the peak at 2922\,$\mu$Hz, which is the central component of an incomplete triplet. We do not identify the two linear combination frequencies since they could be resonant modes \citep{2016A&A...585A..22Z} or pure non-linear effects of the flux perturbation \citep{1995ApJS...96..545B}.

\section{Theoretical results}
\subsection{Asteroseismic models}

\begin{table*}
    \caption{Parameter space explored by the \texttt{WDEC} code during optimization for EPIC~228782059.}
    \centering
    \label{tab:2_table}
    \begin{tabular}{lccccc}
        \hline
        Parameters & Range & Crude & Fine & Sophisticated & Optimal \\
        \hline
        $M_{*}$/$M_{\odot}$                  &[0.550,0. 850]&0.010&0.005&0.005&0.685$\pm$0.003\\
        $T_{\rm eff}$(K)&[20000, 32000]&200&50 &10 &21910$\pm$23\\
        $-$log($M_{\rm env}/M_{\rm *}$)&[2.00,5.00]&1.00&0.03&0.01&2.11$\pm$0.03\\
        $-$log($M_{\rm He}/M_{\rm *}$)&[3.00,6.00]&1.00&0.03&0.01&2.91$\pm$0.02\\
        $X_\mathrm{He}$&[0.00,1.00]&0.20&0.03&0.01&0.59$\pm$0.05\\
        \hline
         Initial $X_\mathrm{O}$ (\%)\\
        \hline
        $h_1$ = 62 &[50,74]&3&1&1&\,52$\pm$7\\
        $^*h_2$ = 68 &[59,77]&3&1&1&\,61$\pm$4\\
        $^*h_3$ = 83 &[77,89]&3&1&1 &\,78$\pm$12\\
        $w_1$ = 38 &[29,44]&3&1&1&\,38$\pm$1\\
        $w_2$ = 51 &[42,57]&3&1&1&\,51$\pm$1\\
        $w_3$ = 9  & [6,12]&3&1&1&\, 9$\pm$1\\
        \hline
    \end{tabular}
    \tablenotetext{}{Notes.\,$M_{\rm env}$ is envelope mass\,; $M_{\rm He}$ is helium mass; { $^*$ The value of $h_{j+1}$ is scaled to $h_{j}$ here $j = 1,2$}}
\end{table*}

\begin{figure}[htpb] 
    \centering 
    \includegraphics[width=\linewidth,scale=1.00]{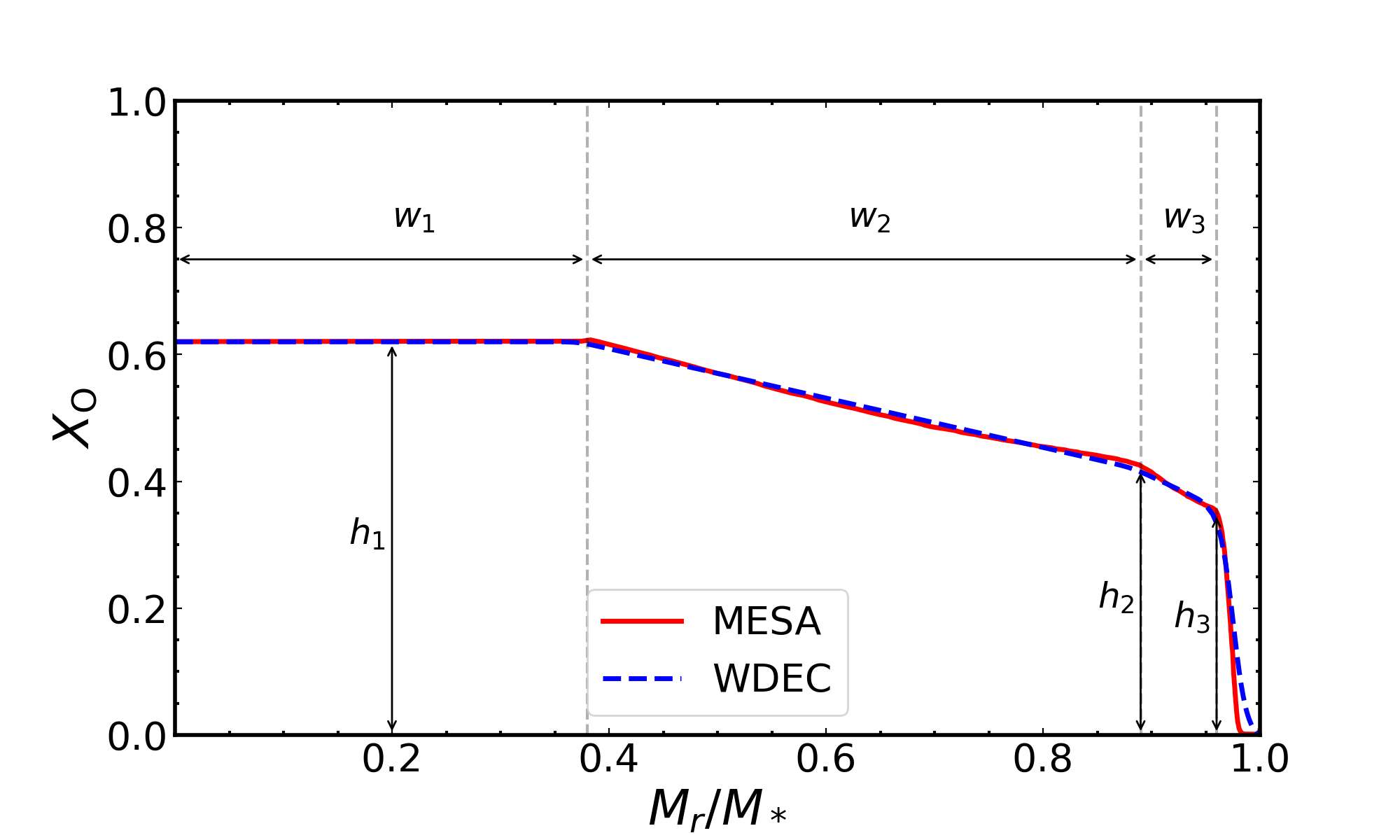}
    \caption{Chemical profiles of oxygen as a function of the mass coordinate.}
    \label{Fig2.eps}
\end{figure}

With the above identifications for the observed modes of $f_{1,0}$, $f_{2,0}$, $f_{3,0}$, $f_{4,0}$, and $f_{5,0}$, those five modes (three $\ell = 1$ and two $\ell = 2$) are used to constrain the theoretical models using the White Dwarf Evolution Code \citep[\texttt{WDEC};][]{1975ApJ...200..306L,2018AJ....155..187B}. The latest version of \texttt{WDEC} incorporates the physical input parameters from the code of Modules for  Experiments in Stellar Astrophysics \citep[\texttt{MESA};][]{2011ApJS..192....3P,2019zndo...2665077P}. The input parameters are crucial to the stellar structure of the build models. There are six most important parameters that can be modified in \texttt{WDEC}: the mass ($M_*$); the effective temperature ($T_\mathrm{eff}$); the location of the base of the envelope ($M_\mathrm{env}$); the location of the helium atmosphere ($M_\mathrm{He}$); the helium abundance in the mixed region ($X_\mathrm{He}$); and the central core oxygen abundance ($X_\mathrm{O}$) that is determined by six parameters ([$h_i, w_i$] and $i=1,2,3$). These parameters describe the shape of the $X_\mathrm{O}$ profile, which is illustrated in Figure\,\ref{Fig2.eps}. 
A monotonous decrease in the oxygen abundance is forced by setting the vertical height parameters to obey $h_1> h_2 > h_3$. While the width parameters $w_1, w_2, w_3$ define the mass fraction of each segment corresponding to $h_1, h_2, h_3$, respectively.

We first evolve a main sequence star with 3.50~$M_{\odot}$ down to a white dwarf with 0.65~$M_{\odot}$ via the \texttt{MESA} code~8118 using default parameters. This process provides a preliminary interior $X_\mathrm{O}$  which was then adjusted slightly to accommodate the white dwarf models that will be calculated for eigenfrequencies by the \texttt{WDEC} code. Figure\,\ref{Fig2.eps} presents the result of oxygen profile obtained by \texttt{MESA} and the initial parameterized $X_\mathrm{O}$ with $h_1 = 0.62$, $h_2 = 0.68h_1$ and $h_3=0.83h_2$. As an alternative, we change the $X_\mathrm{O}$ slightly to better constrain the optimal models, for instance, $h_1 \in [0.50, 0.74]$ as listed in Table\,\ref{tab:2_table}. As suggested by spectroscopy, the $T_\mathrm{eff}$ was set to be in the range of [20,000, 32,000]~K with a rough step of 200 K, to fully cover the entire region of DBV strip. The other parameters were set in a space of a typical DB white dwarf, as listed in Table\,\ref{tab:2_table}.

We then calculated a series of models to reproduce the observed frequencies. The quality of the fitting is evaluated quantitatively by a merit function, defined as:
\begin{equation}
	\chi^{2} = \frac{1}{N} \sum_{i=1}^N (P_{\mathrm{cal}}-P_{\mathrm{obs}})^2,
\end{equation}
where $P_{\mathrm{cal}}$ and $P_{\mathrm{obs}}$ are the calculated and the observed period, respectively. Here $N$ is the number of the independent modes with $m=0$, which is five for EPIC~228782059.

\begin{figure}[htpb] 
    \centering 
    \includegraphics[width=\linewidth,scale=1.00]{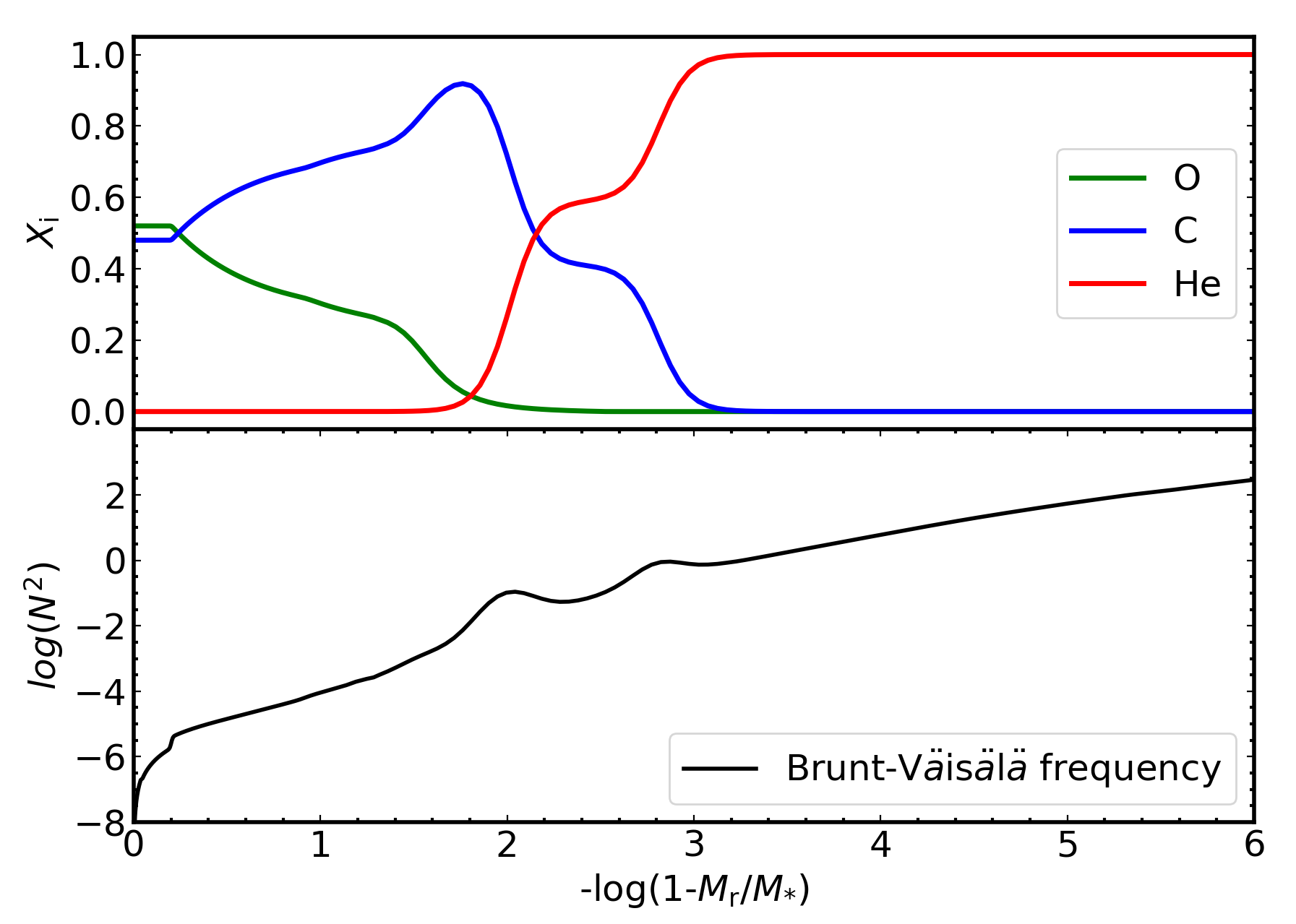}
    \caption{ Top panel: Chemical composition profiles of oxygen, carbon and helium as a function of the outer mass fraction from our seismic model. Bottom panel: The corresponding logarithm of the squared Brunt-V\"{a}is\"{a}l\"{a} frequency }
    \label{fig:figure_add.eps}
\end{figure}

We can quickly locate the good models in the parameter space using large steps, followed by finer steps in a narrower range of the better models. The optimal model was searched using more sophisticated steps. One can see the detailed values of those steps in Table\,\ref{tab:2_table}. The seismic model revealed for EPIC~228782059 has 
$T_\mathrm{eff}=21910\pm23$\,K, roughly consistent with that derived from the {\sl SDSS} spectrum \citep{2015A&A...583A..86K}. The mass is constrained from asteroseismology to be $0.685\pm0.003~M_{\odot}$. The period discrepancy has a value of $\chi = 0.131$\,s between the observed values and theoretical values (see individual periods in Table\,\ref{tab:1_table}). With transformation of luminosity to magnitude, we note that the optimal seismic luminosity is $L=0.028~L_{\odot}$, corresponding to a seismic distance of 343.56\,pc, which is consistent with the result derived from the {\em Gaia} distance $372.38_{-29.34}^{+33.81}$~pc \citep{2021AJ....161..147B}. Figure\,\ref{fig:figure_add.eps} presents the chemical composition profiles of $X_\mathrm{O}$, carbon ($X_\mathrm{C}$) and helium ($X_\mathrm{He}$) as well as  the corresponding Brunt-V\"{a}is\"{a}l\"{a} frequency. We can see small bumps arising at the zones of element transition. However, our models are limited to central C/O profiles and do not incorporate other chemical elements such as $^{22}$Ne. \citet{2021ApJ...910...24C} suggest that the presence of $^{22}$Ne might lead to a systematic offset of pulsation period in the fitting process of specific white dwarfs.

\subsection{Instability strip}

\begin{figure*}
	\center 
	\includegraphics[width=18cm,angle=0]{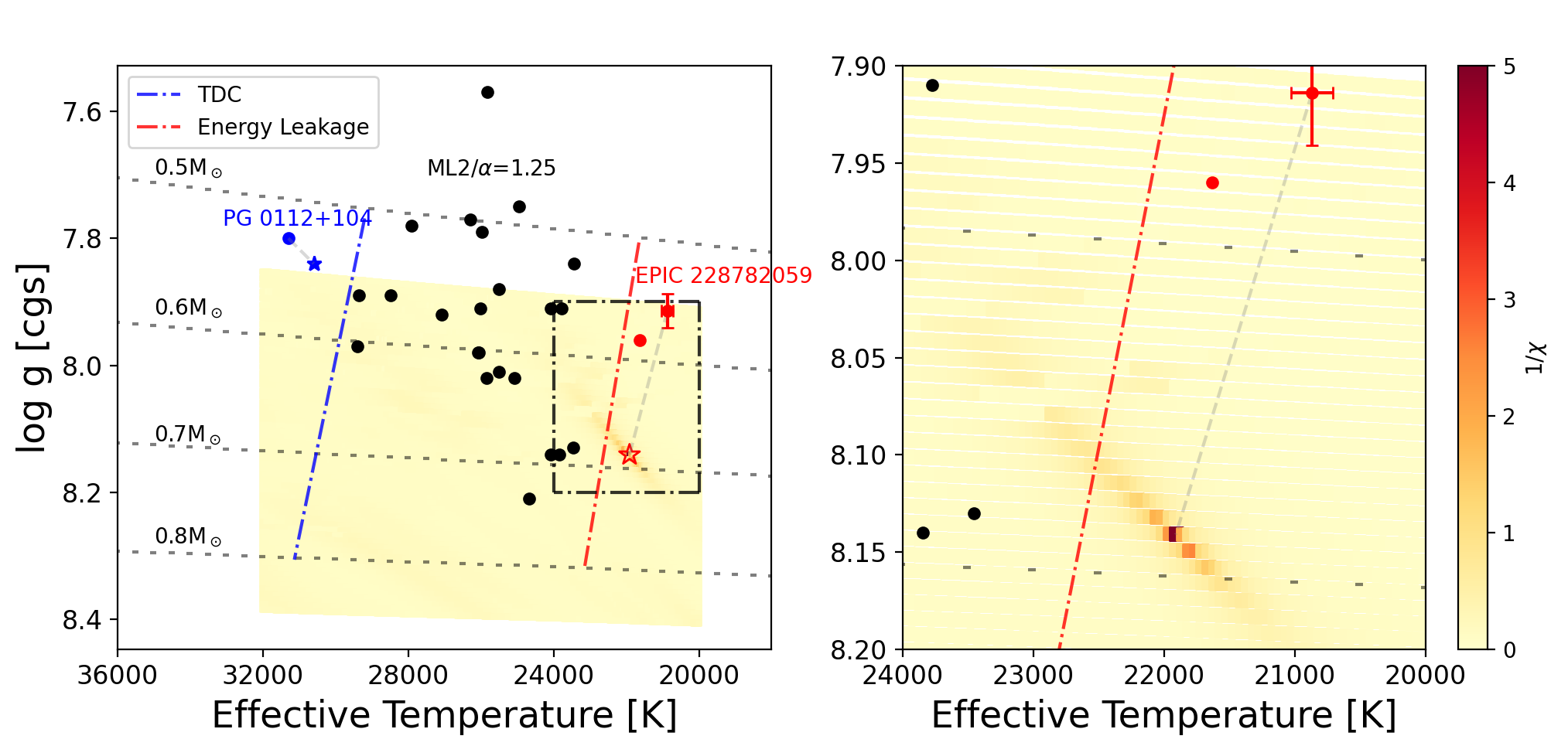}
	\caption{The location of EPIC~228782059, marked as a red square, in the DBV instability strip. All filled circles represent DB variables and the ``star'' symbol is linked to the same star but its parameters were derived from seismic model. The shadow region covers the parameter space  where the \texttt{WDEC} code has been explored. The right panel is the expanded view of the area covered by the dashed box in the left panel. The blue and red dashed lines indicate the two edges that are calculated with different theoretical treatments (see details in text).}
	\label{fig:best_chi_bin_logg_0131.eps}
\end{figure*}

Contrary to its cousin, the hydrogen-atmosphere DAV stars, there are far fewer DBV stars with detected pulsations, and DBVs account for a small fraction of all pulsating white dwarf stars known \citep{2019A&ARv..27....7C}. Thus the determination of the instability strip has attracted less attention for the DBVs. However, the number of DBV stars is still increasing, with many intensive observational efforts \citep[see, e.g.,][]{2018..New..DBVs}.

Given our optimal asteroseismic model, we can locate the position of EPIC~228782059 on the Kiel diagram ($T_\mathrm{eff}$ v.s. $\log g$), with the determination of $\log g$ of 8.14~dex (Figure\,\ref{fig:best_chi_bin_logg_0131.eps}). Along with the other 24 known DBV stars with derived parameters, EPIC~228782059 sits very near the red edge of the DBV instability strip, whereas the other {\em K2}  helium-atmosphere pulsator, PG~0112+104, represents the hottest known DBV \citep{2017ApJ...835..277H}. As an observational constraint to the DBV instability strip, we also adopt the theoretical calculations by \citet{2017ASPC..509..321V}  with several different treatments\footnote{The full time-dependant convection (TDC) and the energy leakage treatment.} in four different cooling tracks of 0.5, 0.6, 0.7 and 0.8$M_\odot$. We note that the evolutionary tracks in Figure\,\ref{fig:best_chi_bin_logg_0131.eps} are slightly different to theirs since the calculations are performed by different codes with the same input parameters, i.e., the mixing length value ML2$/\alpha = 1.25$.

We see that most of the known DBV stars reside well within the theoretical instability strip except for PG~0112+104, the hottest one, as well as EPIC~228782059 and SDSS~J102106.69+082724.8, the two coolest ones, which are important to constrain the location of the blue and red edge, respectively. PG~0112+104 stands about 1000~K hotter than the blue edge by the TDC treatment. However, \citet{2009JPhCS.172a2075C} explored the different values of mixing length to determine the location of the blue edge and found that ML3$/\alpha = 2$ can reproduce a blue edge to fully include PG~0112+104. Detailed calculations give a roughly 1000~K cooler red edge when energy leakage is taken into consideration \citep[see details in][]{2017ASPC..509..321V}. This is favored except that the two cool DBV stars were measured with a large uncertainty in $T_\mathrm{eff}$ and $\log g$.

Strong evidence is provided from the fact that $T_\mathrm{eff}<22{,}000$~K is independently derived both from spectroscopy and asteroseismology using the \texttt{WDEC} code for EPIC~228782059. The seismic $T_\mathrm{eff}$ determination is measured based on a pure helium atmosphere \citep{2015A&A...583A..86K}. The presence of a small fraction of hydrogen may have a large effect on the determination of $T_\mathrm{eff}$, as suggested by \citet{2011ApJ...737...28B}, who claim the DBV instability strip could actually be divided into two regions, named DB and DBA instability strip, taking the hydrogen in atmosphere into account. The theoretical calculations by
\citet{2017ASPC..509..321V} do not find this distinction. The fits of \citet{2015A&A...583A..86K} put an upper limit on the detection of hydrogen in EPIC~228782059 of $\log$~N(H)/N(He) $< -4.2$.

 However, the relative short pulsation periods disclosed here, 228.7-342.2\,s, are comparable to many hot DBVs, for instance KIC\,08626021 and PG~0112+104, where they typically develop shallower convection zones than cooler ones, driving shorter-period pulsations. The SDSS spectrum had also been independently analysed by \citet{2018PASP..130h4203K} who obtained significantly different parameters: $T_{\rm{eff}} =29776\pm258$\,K and $\log g =7.88\pm0.014$\,dex. But this result converts into a distance of $561.3_{-10.2}^{+13.1}$\,pc, indicating that EPIC~228782059 locates farther away from the {\em Gaia} distance.  

To fully establish the observational instability strip of DBV stars, a homogeneous work should be extensively performed to analyze high-quality spectra with serious caution on the hydrogen abundance for known DBVs, including EPIC~228782059, as suggested by \citet{2011ApJ...737...28B}. The determination of stellar parameters ($T_\mathrm{eff}$ and $\log g$) can be used to map the instability strip of DBV stars, which provides constraints on theoretical expectations.

\section{Conclusions}
We report the discovery of 11 independent pulsation modes in the helium-atmosphere pulsating white dwarf (DBV) EPIC~228782059 using more than two months of {\em K2} photometry. The light curves, spanning nearly 70~d, have been extracted with an optimal 6-pixel aperture and corrected to minimize systematic effects from the spacecraft. We detect 11 frequencies and 2 linear combinations above the 4.8$\sigma$ confidential level. From the average frequency spacing of  $\Delta f \approx 4.0$ and 6.8\,$\mu$Hz for $\ell = 1$ and 2 modes, respectively, we measure a rotation period of $34.1\pm0.4$~hr for EPIC~228782059. 

With mode identifications from the rotational splittings, a series of grids were constructed using the \texttt{WDEC} evolutionary code based on the atmospheric parameters from spectrum: $T_{\rm{eff}}=20{,}868\pm160$\,K, $\log g = 7.914 \pm0.027$~dex \citep{2015A&A...583A..86K}. The optimal model of EPIC~228782059 is obtained with a period difference between the theoretical and observed $m=0$ modes of 0.131\,s. Our asteroseismic solution is similar to the values derived from the SDSS spectrum: we find $T_{\rm{eff}}=21{,}910\pm23$~K\ and $M = 0.685\pm0.003~\mathrm{M}_\odot$ from the {\em K2} pulsation modes. If this white dwarf truly has $T_{\rm{eff}}<22{,}000$\,K it would be one of the coolest DBVs known.

The location of EPIC~228782059 on the Kiel diagram can be helpful to test the purity and extent of the DBV instability strip. Our results suggest that EPIC~228782059 is slightly cooler than the red edge predicted by theoretical calculations \citep{2017ASPC..509..321V}. However, our determinations assume a completely pure helium atmosphere; the presence of even trace amounts of hydrogen could lead to a different $T_{\rm{eff}}$. We foresee future analysis of this star using higher-resolution spectroscopy, comparison with other seismic models, consideration of various elements in the core as well.

\section*{Acknowledgements}

We acknowledge the discussions with Valerie Van Grootel and Bell Keaton which are helpful to improve the manuscript. We acknowledge the support from the National Natural Science Foundation of China (NSFC) through grants 11833002, 11903005, 12090040, 12090042 and 11803004. W.Z. is supported by the Fundamental Research Funds for the Central Universities. J.J.H. acknowledges support provided by NASA {\em K2} Cycle 6 Grant 80NSSC19K0162. S.C. is supported by the Agence Nationale de la Recherche (ANR, France) under grant ANR-17-CE31-0018, funding the INSIDE project, and financial support from the Centre National d'Études Spatiales (CNES, France). The authors gratefully acknowledge the {\em Kepler} team and all who have contributed to making this mission possible. Funding for the {\em Kepler} mission is provided by NASA's Science Mission Directorate.

\software WDEC (Bischoff-Kim \& Montgomery 2018), LPCODE (Corsico et al. 2013), Lightkurve (Lightkurve Collaboration et al. 2018), FELIX (Charpinet et al. 2010), MESA (v8118; Paxton et al. 2011; Paxton 2019).

\bibliographystyle{aasjournal}

\end{document}